\RequirePackage{fix-cm}
\documentclass[twocolumn,epjc3]{svjour3}  
\smartqed  
\RequirePackage{graphicx}
\RequirePackage{amsmath}
\RequirePackage{color}
\newcommand{\rhz}[1][m]{\rho_{0, #1}}
\newcommand{\ofrt}{\left( \mathbf{r}, t \right)}
\newcommand{\dqe}{d q^2 + \eta}

\newcommand{\almurh}[1]{\alpha_#1 \mu_#1 \rhz[#1]}

\journalname{Eur. Phys. J. E}
\begin{document}

\title{Non-equilibrium phase separation in mixtures of catalytically active particles: size dispersity and screening effects}



\author{Vincent Ouazan-Reboul\thanksref{addr1}
        \and
        Jaime Agudo-Canalejo\thanksref{addr1} 
        \and
        Ramin Golestanian\thanksref{e3, addr1, addr2}
}


\thankstext{e3}{e-mail: ramin.golestanian@ds.mpg.de}


\institute{Max Planck Institute for Dynamics and Self-Organization, Am Fassberg 17, D-37077, G\"{o}ttingen, Germany\label{addr1}
           \and
           Rudolf Peierls Centre for Theoretical Physics, University of Oxford, OX1 3PU, Oxford, UK  \label{addr2}
}

\date{Received: date / Accepted: date}

\abstractdc{
Biomolecular condensates in cells are often rich in catalytically-active enzymes. This is particularly true in the case of the large enzymatic complexes known as metabolons, which contain different enzymes that participate in the same catalytic pathway. One possible explanation for this self-organization is the combination of the catalytic activity of the enzymes and a chemotactic response to gradients of their substrate, which leads to a substrate-mediated effective interaction between enzymes. These interactions constitute a purely non-equilibrium effect and show exotic features such as non-reciprocity. Here, we analytically study a model describing the phase separation of a mixture of such catalytically-active particles. We show that a Michaelis-Menten-like dependence of the particles' activities manifests itself as a screening of the interactions, and that a mixture of two differently-sized active species can exhibit phase separation with transient oscillations. We also derive a rich stability phase diagram for a mixture of two species with both concentration-dependent activity and size dispersity. This work highlights the variety of possible phase separation behaviours in mixtures of chemically-active particles, which provides an alternative pathway to the passive interactions more commonly associated with phase separation in cells. Our results highlight non-equilibrium organizing principles that can be important for biologically relevant liquid-liquid phase separation.
}

\maketitle

\section{Introduction}
\label{sec:intro}
Enzymes, which are chemically-active proteins that catalyze metabolic reactions, have been found to exhibit non-equilibrium dynamical activity \cite{AgudoCanalejo2018}. 
As part of their biological function, they are also known to self-organize into clusters called metabolons, which contain different enzymes that participate in the same catalytic pathway \cite{Sweetlove2018}.
One possible theoretical explanation for this process is based on the ability of enzymes to chemotax in the presence of gradients of their substrate, which has been
experimentally observed in recent years for a variety of enzymes \cite{Yu2009,Sengupta2013,Dey2014,Zhao2018,Jee2018}. 
The mechanisms underlying enzyme chemotaxis, however, are as of yet still unclear, with diffusiophoresis and substrate-induced changes in enzyme diffusion being possible candidates
\cite{Jee2018,agudo2018phoresis,Tunrayo2019,AgudoCanalejo2020,Feng2020}.
In a recent publication \cite{AgudoCanalejo2019}, it was shown that the interplay between catalytic activity and chemotaxis can lead to effective non-reciprocal interactions \cite{soto14,soto15,Saha2019} between enzyme-like particles, resulting in an active mechanism for the phase separation of such particles.
This active phase separation is distinct from the non-equilibrium phase separation
models that have been more commonly put forward in the cell biological context \cite{weber2019physics}, 
where the interactions between the different components are equilibrium ones,
and the non-equilibrium aspect comes from fuelled chemical reactions that act as a source
or sink of some of the phase-separating components.
In contrast, in the model of Ref.~\cite{AgudoCanalejo2019}, the phase-separating components are conserved, and it is the effective interactions between them that represent an
intrinsically out-of-equilibrium phenomenon. For the particular case of a suspension of a single type of enzymes, the resulting aggregation process was later studied theoretically in more detail in Ref.~\cite{giunta2020cross}. 
An interesting non-biological model system to study these effects is provided by catalyst-coated synthetic colloids \cite{RG2012,Saha2014} and chemically-active droplets \cite{Stark2016}, which have experimentally been shown to form aggregates \emph{via} chemical-mediated effective interactions \cite{pala13,niu17a,yu18,mass18,schm19,meredith2020predator}.

Here, we will generalize the model studied in Ref.~\cite{AgudoCanalejo2019}, by accounting for size polydispersity of 
the catalytically-active particles involved in the mixture, as well as for the the dependence of catalytic activity on the concentration of substrate.
We show that taking into account the dependence on substrate concentration leads to screening effects, which put
a stricter activity threshold for the occurrence of a spatial instability.
Moreover, we show that a mixture of different-sized catalytically-active particles can undergo both local and system-wide self-organization, with
the latter possibly showing oscillatory phenomena.
A model that simultaneously takes into account both of these effects is finally shown to exhibit a rich phase diagram, ranging from non- to partially- to fully-oscillatory.

The paper is organized as follows. In section \ref{sec:model_pres}, we explain the model describing the chemically
active particles, and summarize previous results on the simplest version of this model \cite{AgudoCanalejo2019}.
In section \ref{sec:screen_same_sized}, we reveal a screening effect created by
a dependence of the catalytic activity on the concentration of substrate, and
conclude that this effect leads to an instability threshold and a local (as opposed
to system-wide) instability.
Then, in section \ref{sec:noscreen_diff_sized}, we study the effect of a difference
in the sizes of different particle species, which enters the theory as a difference in their diffusion
coefficient.
We show that, under these conditions, the stability phase diagram of the particle mixture
shows an extended instability region corresponding to a local instability,
and can also exhibit transient oscillations during a system-wide instability.
Finally, in section \ref{sec:full_model}, we consider both screening and size
dispersity effects combined, which leads to a complex stability phase diagram, which includes
fully-, partially-, and non-oscillatory local instabilities.

\section{Linear stability analysis of a chemically-active mixture}
\label{sec:model_pres}
\subsection{Model for chemically-active particles}
\label{sec:evol_eqs}
\begin{figure}[t]
    \centering
    \includegraphics[width=\columnwidth]{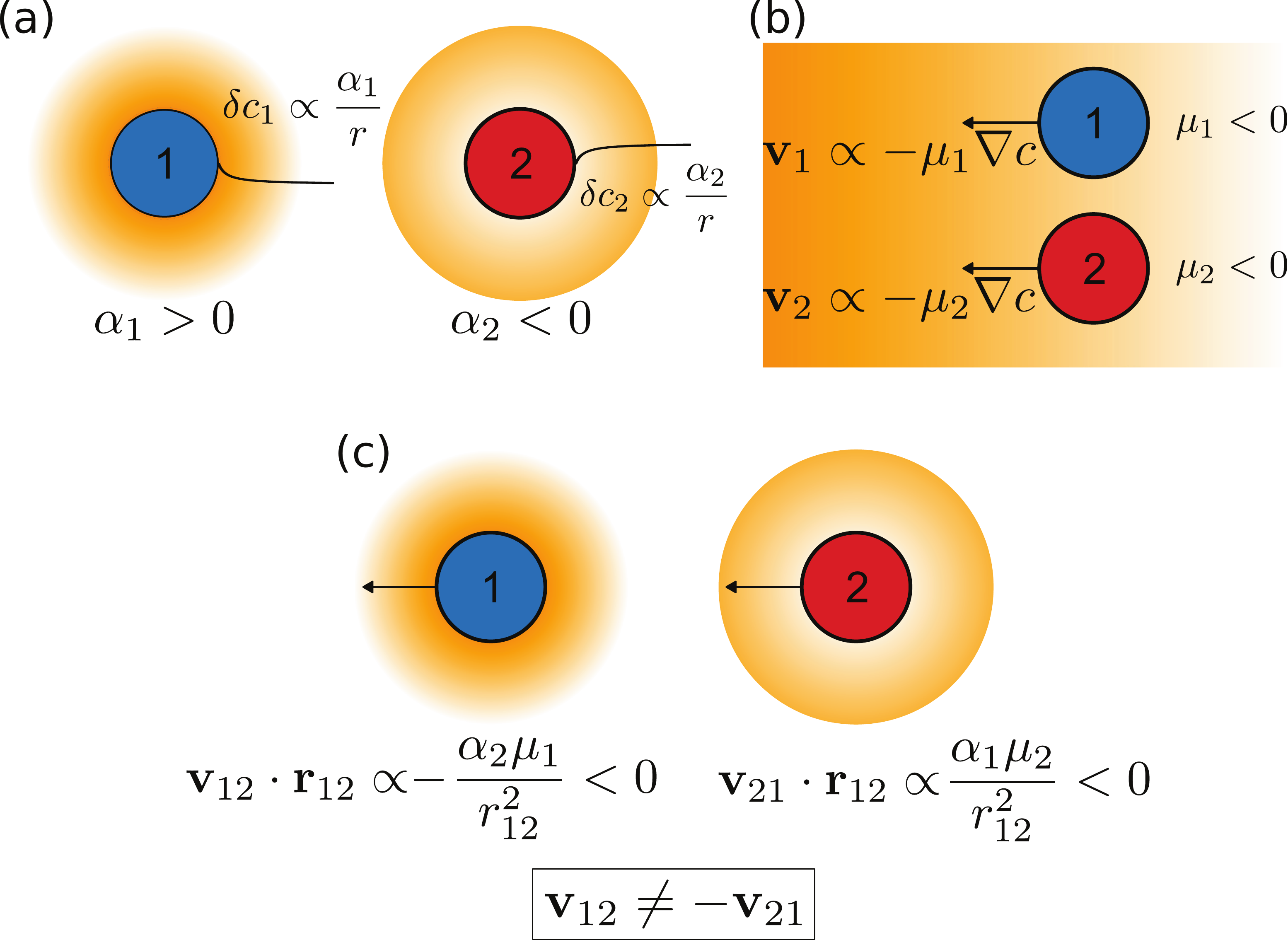}
    \caption{Model for chemically-active particles.
        \textbf{(a)} Chemical activity: particles 1 and 2 respectively produce and
        consume a chemical species (orange), perturbing its concentration profile around them.
        \textbf{(b)} Chemotaxis: the two species develop a velocity in response to
        concentration gradients 
        of the same chemical they act on, in this case towards higher concentrations.
        \textbf{(c)} Particle-particle interactions arising from the combination of 
        these two properties.
        Each particle both perturbs the chemical field and responds to the other's
        perturbation, leading to non-reciprocal interactions  characteristic of active mixtures. 
        In this case, species 1 is attracted by species 2, which is itself repelled by 1, giving rise to a chasing interaction.
    }
    \label{fig:active_parts}
\end{figure}

We study chemically-active particles (for instance, enzymes or catalyst-coated colloids) 
whose chemical activity is characterized by a parameter
$\alpha$, which is the rate at which they consume ($\alpha<0$) or produce ($\alpha>0$)
a given chemical.
If we denote $c$ the concentration of this chemical species, the presence of an isolated active particle creates
a long-ranged perturbation to the concentration field of the chemical, which in steady state goes as
$\delta c \propto \frac{\alpha}{r}$ (Fig.~\ref{fig:active_parts}a).

The considered active particles are also chemotactic: in a concentration gradient
of the chemical they act on, they develop a velocity 
$\mathbf{v} \propto -\mu \nabla c$ (Fig.~\ref{fig:active_parts}b), 
which drives them towards high concentrations 
if $\mu < 0$ (chemotaxis), and low concentrations if $\mu > 0$ (antichemotaxis).
Synthetic colloids can be engineered to be chemotactic, for instance using phoretic
effects \cite{ande89,gole07}. 
Meanwhile, many enzymes have been reported to chemotax in gradients
of their substrate \cite{Yu2009,Sengupta2013,Dey2014,Zhao2018,Jee2018}, with a variety of mechanisms having been proposed to explain the phenomenon
\cite{AgudoCanalejo2018,Jee2018,agudo2018phoresis,Tunrayo2019,AgudoCanalejo2020,Feng2020}.

These two properties give rise to effective particle-particle interactions
mediated by the chemical field,
which takes the form of a velocity developed by particle $i$
in the presence of particle $j$ given by \cite{AgudoCanalejo2019,soto14,soto15}
\begin{equation}
    \label{eq:interaction}
    \begin{aligned} {\mathbf {v}}_{ij} \propto - \frac{\alpha_j \mu_i}{r_{ij}^3} {\mathbf {r}}_{ij}
    \end{aligned}
\end{equation}
with $\mathbf{r}_{ij} = \mathbf{r}_i - \mathbf{r}_j$ the inter-particle distance vector.
Note that, as the perturbation of and the response to the concentration field
obey to different parameters, this interaction is in general non-reciprocal:
$\mathbf{v}_{ji} \neq -\mathbf{v}_{ij}$,
leading for instance to the possibility of chasing interactions (Fig.~\ref{fig:active_parts}c). This non-reciprocity, characteristic of active matter systems \cite{AgudoCanalejo2019,soto14,soto15,Saha2019,ivlev2015statistical,saha20,you2020nonreciprocity,Fruchart2021}, 
can give rise to interesting 
many-body phenomena, which we will study here.

\subsection{Linear stability analysis}
\label{sec:lin_stab}

We wish to study the ability of a mixture of these active particles to self-organize.
To do so, we consider $M$ species of particles, each with an activity $\alpha_m$ and a
mobility $\mu_m$, and described by a concentration field $\rho_m \ofrt$. All the species act on the same chemical field, which we may refer as the messenger 
chemical.
The active species concentrations evolve according to the Smoluchowski equation:
\begin{equation}
    \label{eq:evol_rho}
    \partial_t \rho_m (\mathbf{r}, t)= \nabla \cdot
    \left[
      D_m \nabla \rho_m
      + \mu_m (\nabla c\ofrt) \rho_{m} \ofrt
    \right]
\end{equation}
with $D_m$ the diffusion coefficient of species $m$ and $c \ofrt$ the concentration
of the chemical.

The concentration of the chemical, meanwhile, obeys a reaction-diffusion equation:
\begin{equation}
    \label{eq:evol_c}
    \partial_t c \ofrt =
    d \nabla^2 c +
    \sum_{m=1}^{M}\Bigl(
      \alpha_{m}(c) \rho_{m}\ofrt
    \Bigr)
\end{equation}
where we allow for the activity of the active particles to be a function of the chemical 
concentration, and with $d$ the  diffusion coefficient of the chemical.

We perform a linear stability analysis by considering perturbations around a spatially-homogeneous steady state, writing:
$\rho \ofrt = \rhz + \delta \rho \ofrt$ and $c \ofrt = c_H (t) + \delta c \ofrt$
with $c_H(t)$ the (potentially time dependent) homogeneous concentration of the messenger chemical. Indeed, this concentration may be time-dependent in cases where the stationarity condition \\$\sum_m \alpha_m(c_H) \rhz[m]=0$ cannot be satisfied, as may occur in systems with non-zero net catalytic activity such as e.g.~producer-only or consumer-only mixtures.

We then expand \eqref{eq:evol_rho} and \eqref{eq:evol_c} to the first order in the perturbations, while
also performing a quasi-static approximation $\partial_t \delta c \ofrt \simeq 0$ in \eqref{eq:evol_c}.
This approximation corresponds to the assumption that the chemical diffuses over timescales much shorter than those associated with the motion of
the active particles, both through diffusion and chemotaxis; as well as with the changes in the overall chemical concentration in mixtures with net catalytic activity.
We also expand the activities to the first order in concentration:
$\alpha_m(c) \simeq \alpha_m ( c_H(t) ) + (\partial_c \alpha_m)|_{c_H} \delta c \ofrt$,
approximating for instance a Michaelis-Menten-like dependence
on the concentration $c$ for the activities.

Note that, in systems with net catalytic activity,
the parameters
$\alpha_m \equiv \alpha_m(c_H(t))$ and 
$(\partial \alpha_m) \equiv (\partial_c \alpha_m)|_{c_H(t)}$
have an implicit time dependence.
Depending on the sign of the total activity $\sum_m \alpha_m \rhz[m]$, the system either homogeneously consumes or
produces the messenger chemical, leading to activity parameters that evolve in time. Only in the special case $\sum_m \alpha_m \rhz[m]=0$ we find a ``neutral'' mixture with no net production or consumption of the chemical.
As we only care about the stability of the system in a given homogeneous state, we will ignore this time dependence in the following.
The time dependence can be brought back into the picture \emph{a posteriori}, for a chosen functional dependence $\alpha_m(c)$,
by considering the trajectories that such a system would describe in parameter space over time.

We look for solutions of the form:
\begin{equation}
    \label{eq:fourier_laplace_sols}
    \begin{aligned}
    &\delta\rho_{m}\ofrt = \sum\limits_{\mathbf{q}, \lambda}
    \delta\rho_{m, \mathbf{q}, \lambda} \ e^{\lambda t + i \mathbf{q\cdot r}}\\
    &\delta c\ofrt = \sum\limits_{\mathbf{q}, \lambda}
    \delta c_{\mathbf{q}, \lambda} \ e^{\lambda t + i \mathbf{q\cdot r}}
    \end{aligned}
\end{equation}
where the $\mathbf{q}, \lambda$ indices will be omitted in what follows, for readability.
By plugging these expressions into the linearized evolution equations, 
we find the eigenvalue problem:
\begin{equation}
  \label{eq:eig_eq}
  \lambda \delta \rho_m = 
  - \frac{q^2}{\dqe}
  \sum\limits_{n=1}^M 
  [\alpha_n \mu_m \rhz[m] + D_m (\dqe) \delta_{mn}] \delta\rho_n
\end{equation}
with $\eta \equiv -\sum\limits_m (\partial \alpha_m) \rhz[m]$ a screening parameter, that is present only when the activities are concentration-dependent.
We note that this screening parameter is generally positive. Indeed, by analogy with Michaelis-Menten kinetics,
the activity of a producer does not depend on the concentration of its product,
and thus $\partial \alpha_m \equiv 0$ when $\alpha_m >0$;
while the activity of a consumer increases with substrate concentration,
and thus $\partial \alpha_m <0$ when $\alpha_m <0$.

Note also that screening may arise in a different way, if we consider that the chemical may undergo spontaneous decay. Such a situation can be taken into account by adding a term $- \kappa c$
in the right-hand side of \eqref{eq:evol_c}, in which case one finds that the screening parameter is rescaled to
$\eta \rightarrow \eta + \kappa$.

Equation~\eqref{eq:eig_eq} features the growth rate $\lambda$ of a given mode as the eigenvalue, 
whose sign will inform us about the stability of the system. 
If at least one eigenvalue is positive, the homogeneous state is unstable and 
the system shows spatial self-organization, typically into dense clusters as seen in particle-based Brownian dynamics simulations of the system \cite{AgudoCanalejo2019}.

At the onset of such an instability, the eigenvector components $\delta \rho_m$ inform us about the stoichiometry 
of the growing perturbation, 
that is, which species tend to aggregate together (and in which proportion),
and which species tend to separate.

\subsection{Simplest case: similarly-sized species without screening}
\label{sec:prl_case}

\begin{figure*}
    \includegraphics[width=\linewidth]{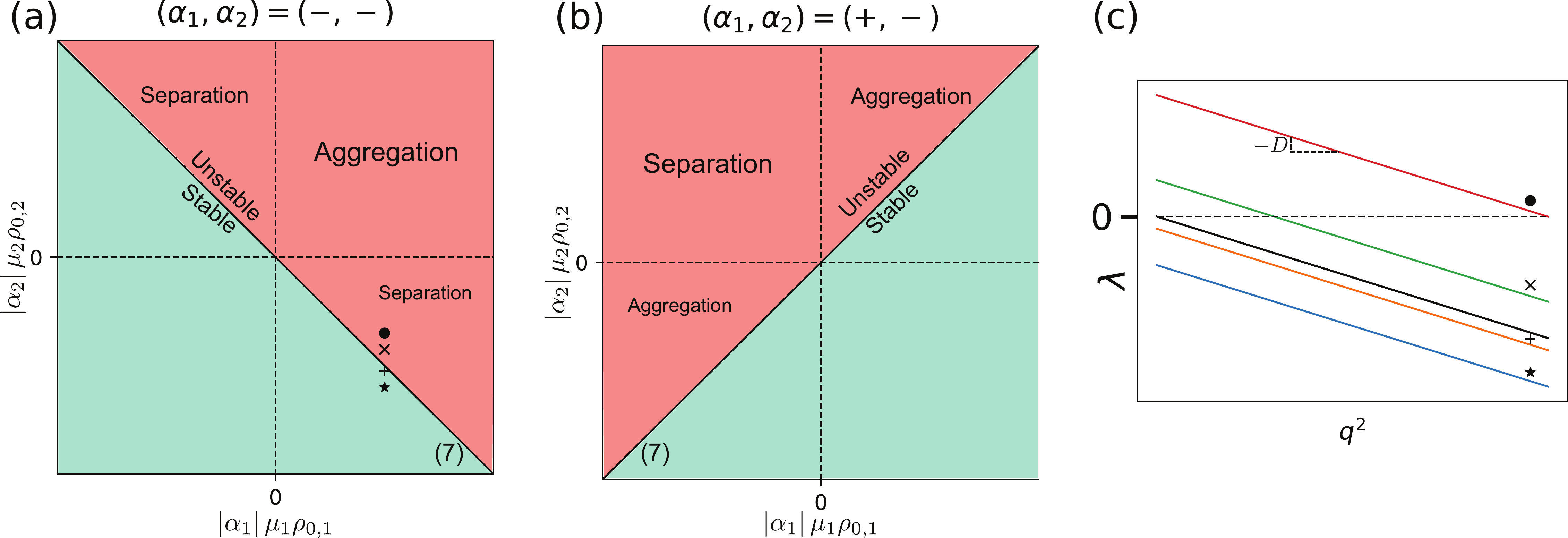}
    \caption{Behaviour of a mixture of two same-sized species with concentration-independent activities \cite{AgudoCanalejo2019}.
        \textbf{(a)} Phase diagrams for two consumers. 
        \textbf{(b)} Phase diagram for one producer, one consumer.
        In (a) and (b), numbers in parentheses refer to the corresponding equations in the text.
        \textbf{(c)} Selected eigenvalue plots as a function of the squared wave vector $q^2$. Coloured lines correspond to the upper eigenvalues
        of the phase diagram points marked in (a). 
        Black line corresponds to the lower eigenvalue, shared by all points in the phase diagram.
    }
    \label{fig:prl_phase_diag}
\end{figure*}

We summarize here the result of the stability analysis for a particularly simple case which was previously 
studied in Ref.~\cite{AgudoCanalejo2019}. If we consider species with concentration-independent activities ($\eta = 0$) and equal sizes 
($D_1 = D_2 = \cdots = D_M = D$), \eqref{eq:eig_eq} reduces to an eigenproblem involving a rank
one matrix, with $M-1$ degenerate eigenvalues $\lambda_-$ 
and one unique eigenvalue $\lambda_+$:
\begin{equation}
    \label{eq:eig_prl}
    \begin{aligned}
    \lambda_- = -D q^2 \\
    \lambda_+ = -D q^2 - \sum\limits_{m=1}^M 
    \frac{\alpha_m \mu_m \rhz[m]}{d}
    \end{aligned}
\end{equation}
Of the two, only $\lambda_+$ can be positive, according to the criterion:
\begin{equation}
    \label{eq:prl_crit}
    \sum_{m=1}^M \alpha_m \mu_m \rhz[m] < 0 
\end{equation}
corresponding to the mixture of active particles being overall self-attractive.
Notice that the only requirement is for the overall sum to be negative,
implying that any arbitrarily small amount of attraction is sufficient to trigger an instability.
This is a consequence of the long-ranged, unscreened nature of the interactions.

When condition \eqref{eq:prl_crit} is satisfied,
the $q^2=0$ mode is the fastest-growing one, and the instability is therefore always system-wide.
The corresponding eigenvector is:
\begin{equation}
    \label{eq:prl_stoich}
    (\delta \rho_1, \delta \rho_2, ...,\delta \rho_M) =
    (\mu_1 \rho_{0,1}, \mu_2 \rho_{0,2}, ..., \mu_M \rho_{0,M})
\end{equation}
The stoichiometry at instability onset is then determined by the mobilities,
independently of the activities.
In particular, species with equal sign of the mobility tend to aggregate together,
whereas those with opposite sign tend to separate.

The behaviour of a two-species mixture can be captured in a two-dimensional phase diagram,
plotted in ($|\alpha_i| \mu_i \rhz[i]$) coordinates for given signs of the activities, i.e.~independently for mixtures of producers and consumers, or for mixtures of two consumers
(Fig.~\ref{fig:prl_phase_diag}).

In the following, we will show that accounting for screening effects due to concentration-dependent activities 
as well as for different-sized particles leads to significant departures from this simple behaviour,
including the existence of a minimum activity threshold for an instability to occur,
and the possibility of oscillatory instabilities.

\section{Screening-induced stability threshold}
\label{sec:screen_same_sized}

\subsection{Arbitrary number of species}
\label{sec:screen_same_sized_general}

\begin{figure*}[hbtp]
    \centering
    \includegraphics[width=\linewidth]{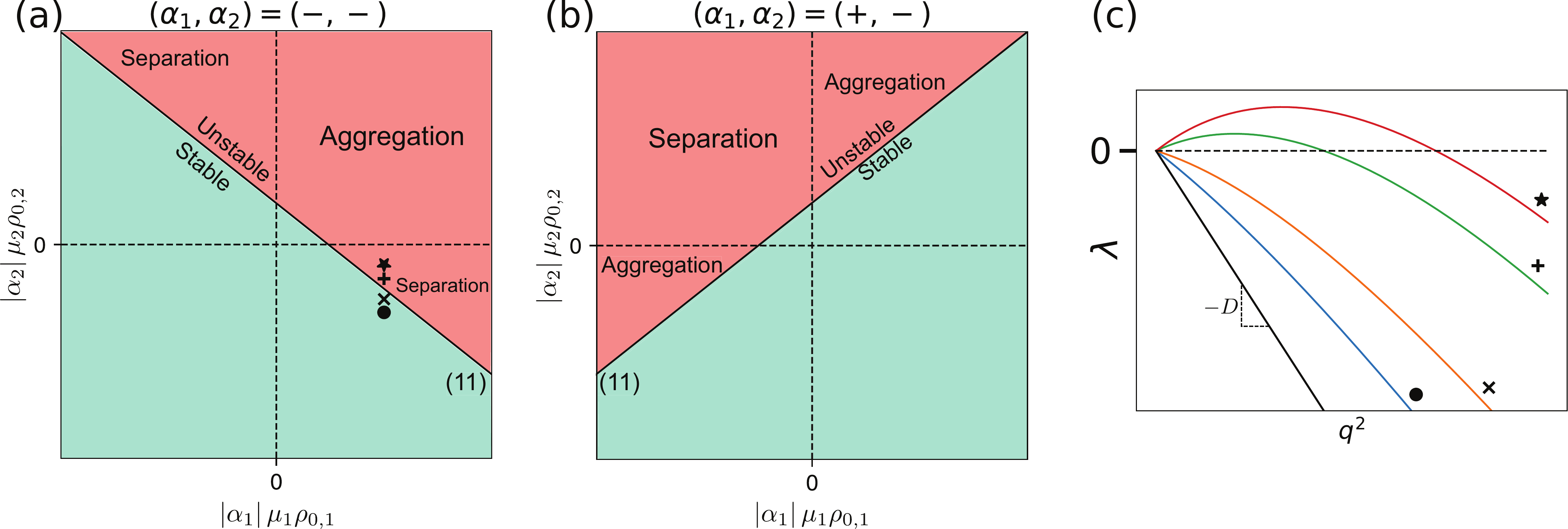}
    \caption{Behaviour of a mixture of two same-sized species with concentration-dependent activities.
        \textbf{(a)} Phase diagram for two consumers.
        \textbf{(b)} Phase diagram for one producer, one consumer.
        In (a) and (b), numbers in parentheses refer to the corresponding equations in the text.
        \textbf{(c)} Eigenvalue plots as a function of the squared wave vector $q^2$.
        Coloured lines correspond to the upper eigenvalue, taken at several locations in (a)
        Black line represents the lower eigenvalue, which does not depend on phase space
    location.}
    \label{fig:phase_diag_same_sized}
\end{figure*}

In the presence of screening ($\eta>0$), but for identically-sized particles $(D_1 = D_2 = ... = D_M = D)$,
the eigenvalue problem \eqref{eq:eig_eq} becomes:
\begin{equation}
    \label{eq:same_sized_lsa}
    \lambda \delta \rho_m = 
    - \frac{q^2}{\dqe}
    \sum\limits_{n=1}^M 
    [\alpha_n \mu_m \rhz[m] + D (\dqe) \delta_{mn}] \delta\rho_n
\end{equation}
which, as in the case described in section \ref{sec:prl_case}, can be reduced to a rank one matrix eigenvalue problem,
with eigenvalues:
\begin{equation}
    \label{eq:same_sized_eig}
    \begin{aligned}
        \lambda_-(q^2) = -D q^2 \\
        \lambda_+(q^2) = 
        -D q^2 - \frac{q^2}{\dqe}
        \sum_{m=1}^M \alpha_m \mu_m \rhz[m]
\end{aligned}
\end{equation}
Once again, only $\lambda_+$ can positive, but this time under the condition:
\begin{equation}
    \label{eq:same_sized_crit}
    \sum_{m=1}^M \alpha_m \mu_m \rhz[m]
    < - \eta D
\end{equation}
This instability criterion corresponds to a stricter version of \eqref{eq:prl_crit},
with the activity dependence on concentration appearing as a screening term.
As a consequence, there is now a threshold value of overall self-attraction required for an instability to occur.

An intuitive way of understanding the more stringent instability criterion is as arising from
a feedback effect affecting consumer species, which are the only ones 
contributing to $\eta$. 
Indeed, these are self-attracting only if they verify $\mu > 0$, i.e.~if they are antichemotactic.
This implies that, in the context of self-attraction, these particles migrate towards
zones of lower chemical concentration, which in turn lowers their activity, and thus their self-attraction. In the self-repelling case, the opposite happens, with a positive feedback on the self-interaction which amplifies the inter-particle repulsion as particles get further away from each other.

Another key difference to the case without screening is that the unstable eigenvalue now 
has a non-monotonic dependence on the wave number $q^2$.
Indeed, we now find $\lambda_+(q^2=0) =0$ always, and the eigenvalue is maximum at
\begin{equation}
    \label{eq:same_size_q2_max}
    q^2 = d^{-1} \left(
        \sqrt{\frac{- \eta \sum_m \almurh{m}}{D}} - \eta
    \right)
\end{equation}
which gives a finite wave length to the fastest-growing perturbations.
With regards to the stoichiometry of the instability, we will show later that the sign of the eigenvector components is still determined by the sign of the mobilities, as before.

\subsection{Two species: phase diagram}
\label{sec:same_sized_twospec}

Figure~\ref{fig:phase_diag_same_sized} shows the phase diagram for a mixture of two
similarly-sized particles with screening.
Comparing it to Fig.~\ref{fig:prl_phase_diag}, we see that the instability line is shifted,
corresponding to the screening-induced instability threshold.
Moreover, the eigenvalue plots also highlight the fact that, while the lower eigenvalues show the same behaviour in both cases, in the screened case, the upper eigenvalue
is zero at $q^2=0$ and goes through a maximum at a finite $q^2$, while in the 
unscreened case, it monotonically decreases from a non-zero value at $q^2=0$.

\section{Differently sized particles: local instability and oscillations}
\label{sec:noscreen_diff_sized}

\subsection{Macroscopic and local instabilities}
\label{sec:noscreen_real}

We now turn to the case of concentration-independent activity ($\eta=0$), but differently-sized particle species.
The eigenvalue problem \eqref{eq:eig_eq} becomes:
\begin{equation}
    \label{eq:eig_problem_noscreen}
    \lambda \delta \rho_m = 
    - 
    \sum\limits_{n=1}^M 
    \left[\frac{\alpha_n \mu_m \rhz[m]}{d} + D_m  q^2 \delta_{mn}
    \right] \delta\rho_n
\end{equation}
involving an arbitrary matrix, now that the diffusion coefficients $D_m$ are species-dependent.
The problem is then intractable in general, and we turn to the two-species case,
which is solvable analytically.
From here on, we choose the convention $D_1 > D_2$ without loss of generality.

Solving for $\lambda$, we find the eigenvalues:
\begin{equation}
    \label{eq:noscreen_eigs}
    \begin{aligned}
        \lambda_\pm(q^2) = 
    &-\frac{1}{2d} \left(
        \gamma_1 + \gamma_2 + (D_1 + D_2) d q^2 
    \right) \\
    &\pm \frac{1}{2 d} \sqrt{
        \left[ \gamma_1 - \gamma_2 + (D_1 - D_2) d q^2 \right] ^2
        + 4 \gamma_1 \gamma_2
    }
    \end{aligned}
\end{equation}
with $\gamma_m = \alpha_m \mu_m \rhz[m]$ the self-interaction of species $m$.
The instability conditions can be obtained by developing the eigenvalues to the first order in $q^2$:
\begin{equation}
    \label{eq:noscreen_eigs_approx}
    \begin{aligned}
        & \lambda_+ (q^2)= -\frac{\gamma_1 D_2 + \gamma_2 D_1}{\gamma_1 + \gamma_2} q^2 + O(q^4)\\
        & \lambda_- (q^2)= -\frac{\gamma_1 + \gamma_2}{d} - 
        \frac{\gamma_1 D_1 + \gamma_2 D_2}{\gamma_1 + \gamma_2} q^2 + O(q^4)
    \end{aligned}
\end{equation}
$\lambda_-$ is unstable when $\gamma_1 + \gamma_2 \leq 0$, or equivalently
\begin{equation}
    \label{eq:instab_no_screen}
    \almurh{1} + \almurh{2} \leq 0
\end{equation}
which coincides with \eqref{eq:prl_crit}, and leads to a system-wide instability (maximum at $q^2=0$).
However, even when $\lambda_-$ is negative, the system can still be unstable, as
$\lambda_+$ can have a positive initial slope when the less strict condition
$\gamma_2 \leq - \frac{D_2}{D_1} \gamma_1$, which we can write as:
\begin{equation}
    \label{eq:ext_instab_line_noscreen}
    \alpha_2 \mu_2 \rhz[2] \leq - \frac{D_2}{D_1} \alpha_1 \mu_1 \rhz[1]
\end{equation}
is verified.
In this case, the instability is only at finite wavelengths as in the screened case,
with $\lambda_+(q^2=0)=0$ and maximum $\lambda_+$ at a finite value of $q^2$.

There is therefore a wider range of conditions under which a mixture can become unstable 
if the particles are differently-sized,
with the caveat that this extended range only leads to a finite wave length instability 
rather than a system-wide one.

\subsection{Transient oscillations}
\label{sec:noscreen_complex}

We can also extract the range of parameters for which the two eigenvalues in \eqref{eq:noscreen_eigs}
become a complex conjugate pair, which results in the condition $\gamma_2 \geq - (D_1/D_2)^2 \gamma_1$,
or equivalently
\begin{equation}
    \label{eq:noscreen_comp_unst_cond}
\almurh{2} \geq -\left(\frac{D_2}{D_1}\right)^2  \almurh{1} 
\end{equation}
where the real parts are positive for a finite range of wavevectors whenever \eqref{eq:instab_no_screen} is satisfied.
The different particle sizes can thus lead to oscillatory instabilities for a finite range of perturbation wave lengths.
Note, however, that the most unstable wave length (eigenvalue with largest real part) still always corresponds to a real eigenvalue,
suggesting that any oscillatory phenomena will be at most transient.

\begin{figure*}
    \centering
    \includegraphics[width=0.8\linewidth]{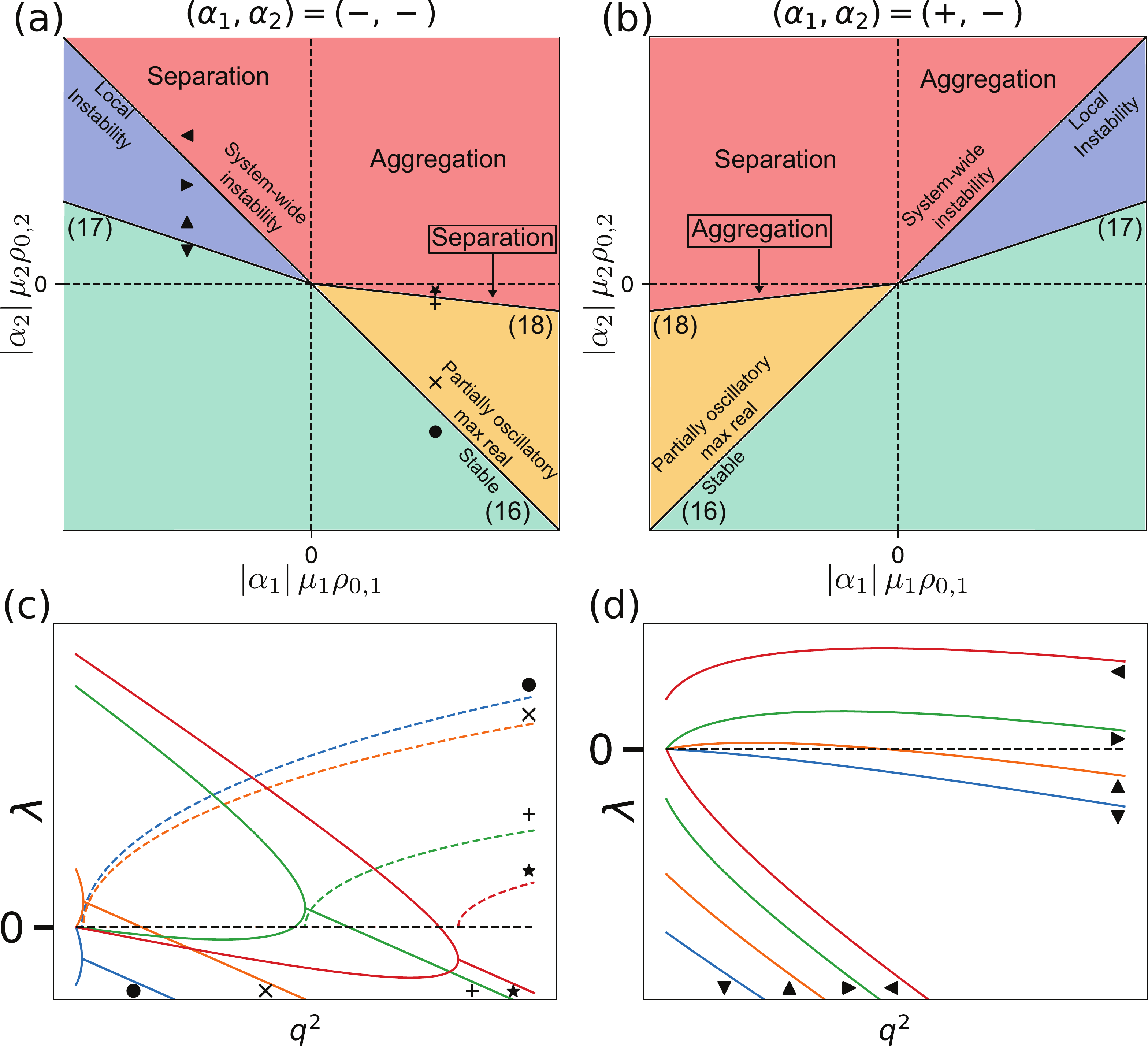}
    \caption{Behaviour of a mixture of two differently-sized species with concentration-independent
        activities.
        \textbf{(a)} Phase diagram for two consumers.
        \textbf{(b)} Phase diagram for one producer, one consumer.
        In (a) and (b), numbers in parentheses refer to the corresponding equations in the text.
        \textbf{(c)} Eigenvalue plots along the transition from stable (blue) to partially oscillatory (orange, green) to real unstable (red).
         Full lines correspond to the eigenvalue real parts, dashed lines to the imaginary part.
         \textbf{(d)} Eigenvalue plots along the  transition from stable (blue) to local (orange, green) and then to macroscopic (red) instability.}
         \label{fig:noscreen_phase_diag}
\end{figure*}

The overall behaviour of the system is summed up in the phase diagrams and eigenvalue plots 
of Fig.~\ref{fig:noscreen_phase_diag}.
Note some marked differences with the cases of sections~\ref{sec:same_sized_twospec} and \ref{sec:prl_case},
most importantly the apparition of  a variety of unstable regions with distinct behaviours.

If $\gamma_1 \geq 0,~\gamma_2 \leq 0$ (lower-right quadrant in Fig.~\ref{fig:noscreen_phase_diag}a, upper-right in~\ref{fig:noscreen_phase_diag}b),
$\lambda_+$ shows similar behaviour to the screened case, being zero at $q^2=0$ with a maximum at finite $q^2$, while $\lambda_-$ has a non-null value at $q^2=0$
(all lines of Fig.~\ref{fig:noscreen_phase_diag}d).
If \eqref{eq:ext_instab_line_noscreen} is verified,
the situation is similar to Fig.~\ref{fig:phase_diag_same_sized}: 
$\lambda_+$ is the unstable eigenvalue, and leads to a local instability (up- and right-pointing triangles on Fig.~\ref{fig:noscreen_phase_diag}d).
However, if \eqref{eq:instab_no_screen} is verified, then $\lambda_-$ becomes positive
and has a positive value at $q^2=0$, leading to a situation similar to
the one in Fig~\ref{fig:prl_phase_diag}, with one key difference:
the positive eigenvalue is non-monotonic, having a maximum at a non-zero wave vector (left-pointing triangle in Fig.~\ref{fig:noscreen_phase_diag}d).
The system should then show an initial, local instability phenomenon followed
by system-wide self-organization.
If $\gamma_1 < 0$ (right and left halves on Fig.~\ref{fig:noscreen_phase_diag}a and b respectively),
the different-sized species mixture shows an entirely new behaviour,
with the eigenvalue being real from $q^2 = 0$ to a finite wave vector, then complex (all lines on Fig.~\ref{fig:noscreen_phase_diag}c).
In the region where the eigenvalues are real, the behavior of the largest one
is similar to Fig.~\ref{fig:prl_phase_diag}, being non-null at $q^2=0$ and
monotonically decreasing.
This behaviour carries over to the real part of the complex eigenvalue, which decreases
monotonically as well, implying that the instability will always be system-wide with the $q^2=0$ mode dominating.
We distinguish between the non- and partly-oscillatory sections of the phase diagram 
by considering
whether or not there exists a region where the eigenvalue is complex with a positive
real part (star label in Fig.~\ref{fig:noscreen_phase_diag}d is non-oscillatory, plus and cross labels are oscillatory).
Finally, the stoichiometry sign for non-oscillatory instabilities is the same
as in section~\ref{sec:prl_case}, as will be shown in the next section.

\section{Variety of behaviours for differently-sized species with screened
    interactions}
\label{sec:full_model}

\subsection{Local instability}
\label{sec:full_model_real}

Finally, we turn to the most general version of the eigenvalue problem \eqref{eq:eig_eq}.
Once again, it is analytically intractable for an arbitrary species number $M$,
and we turn to the $M=2$ case. The solution to \eqref{eq:eig_eq} writes:
\begin{equation}
    \label{eq:full_model_real_eig}
        \begin{aligned}
        \lambda_{\pm} =
        &\frac{1}{2} \frac{q^{2}}{dq^{2}+\eta}
             \Bigg\{
            -(\gamma_{1} + \gamma_{2}) - (D_{1} + D_{2})(dq^{2}+\eta) \\
            &\pm \sqrt{
                \left[ \gamma_1 - \gamma_2 + (D_1 - D_2)(\dqe) \right] ^2
                + 4 \gamma_1 \gamma_2
            }
        \Bigg\}
          \end{aligned}
\end{equation}
We can look for an instability condition either by proceeding as in~\ref{sec:noscreen_real}
and developing the expressions to first order, or by calculating the range of squared
wave vectors for which the eigenvalues are positive.
Imposing $\lambda_+ \geq 0$ leads to two possible conditions, which correspond to two
distinct instability lines.
Since only one of the conditions needs to be satisfied in order for the instability to occur,
we only need to consider the largest of the two instability lines in a given region.
The two lines intersect at the branching point given by
\begin{equation}
    \label{eq:fork_point}
    \left( \left( \almurh{1} \right)_B, \left( \almurh{2} \right)_B \right) 
    = 
    \left( \frac{-\eta D_1^2}{D_1-D_2}, \frac{\eta D_2^2}{D_1-D_2} \right)
\end{equation}
leading to the two instability conditions
\begin{equation}
  \label{eq:instab_full_model_L}
    \almurh{2} \leq - \almurh{1} - \eta(D_{1} + D_{2}) 
    \quad
\end{equation}
for $\almurh{1} \leq (\alpha_1 \mu_1 \rhz[1])_B$, and
\begin{equation}
    \label{eq:instab_full_model_R}
    \almurh{2} \leq
    - \frac{D_{2}}{D_{1}}\almurh{1} - \eta D_{2}
\end{equation}
for $\almurh{1} \geq (\almurh{1})_B$. These two lines recover both the screening-induced shift of the instability line, as well as the extension of
the instability region caused by the different species sizes. However, as opposed to the eigenvalues \eqref{eq:noscreen_eigs}, here both eigenvalues are
null at $q^2=0$: the eigenvalue has the same behaviour in the extended instability
region as in the standard one, and the instability is always local.

\begin{figure*}[hbtp]
    \includegraphics[width=\linewidth]{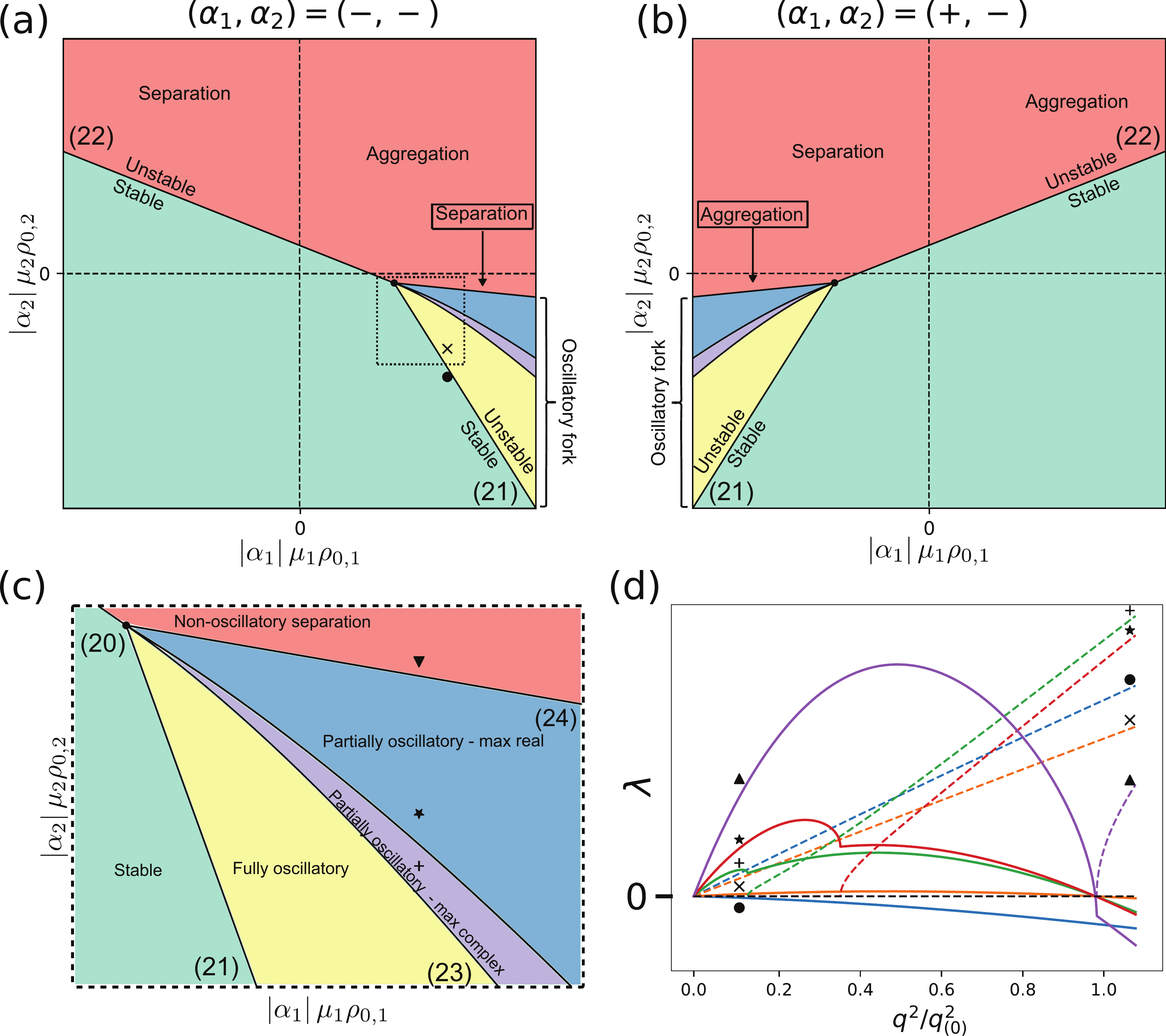}
    \caption{Behavior of a mixture of two differently-sized species with concentration-dependent
        activities.
        \textbf{(a)} Phase diagram for two consumers.
        \textbf{(b)} Phase diagram for one producer, one consumer. 
        \textbf{(c)} Structure of the phase diagram near the 
        branching point for two consumers
        , corresponding to the zoomed-in dashed square in (a).
        In (a), (b) and (c), numbers in parentheses refer to the corresponding equations in the text.
        \textbf{(d)} Eigenvalue plots (see (a) and (c) for marker locations in phase space) along the transition from stable (blue) to fully oscillatory (orange) to partially oscillatory with dominant oscillatory modes (green) and then dominant non-oscillatory modes (red), and finally to non-oscillatory (purple).
        Only the upper eigenvalue is plotted, for readability.
        Full lines correspond to the eigenvalue real parts, dashed lines to the imaginary part.
        Wave vectors are normalized by the largest unstable wave vector $q_{(0)}^2$, if applicable.
    }
    \label{fig:phase_diag_full}
\end{figure*}

\subsection{Partial and fully oscillatory instabilities}
\label{sec:full_model_complex}

Proceeding similarly to section~\ref{sec:noscreen_complex}, we now look for complex eigenvalues.
In phase space, the parameters allowing for complex solutions correspond to a ``fork''
which opens in the instability line for $\gamma_1 \leq \gamma_{1,B}$, from the branching point defined in
\eqref{eq:fork_point}.
By comparing the maximal unstable wavevector to the range of wavevectors for which the eigenvalues
are complex, we find a variety of possible behaviours. When the condition
\begin{equation}
    \label{eq:full_osc}
    \almurh{2} > \left(\sqrt{-\almurh{1}} - \sqrt{\eta (D_1 - D_2)}\right)^2
 \end{equation}
is verified, the full range of wavevectors that are unstable (eigenvalue with positive real part) have complex conjugate eigenvalues, and we term this a ``fully oscillatory'' instability.
Another kind of instability, which we call ``partially oscillatory'', is observed if
condition~\eqref{eq:full_osc} is not satisfied and instead:
\begin{equation}
    \label{eq:part_osc}
    \almurh{2} > 
    - \left(\frac{D_2}{D_1}\right)^2 \almurh{1}
\end{equation}
in which case
the there are two ranges of unstable wave vectors, one with real and one with complex conjugate eigenvalues, each of which features a local
maximum of the real part of the eigenvalue.
Far away from the lower bound given by \eqref{eq:part_osc}, the fastest growing mode is still complex,
and so the instability process should still be mainly oscillatory.
Below a line which can be calculated numerically, as we approach the lower bound given by \eqref{eq:part_osc}, the two maxima cross over, and the global
maximum of the real part occurs for a real eigenvalue, so that the non-oscillatory instability should dominate.
Finally, if the condition \eqref{eq:part_osc} is not satisfied,
then all unstable modes are real and the instability should display no oscillations whatsoever.

The behaviour of the system is summarized in Fig.~\ref{fig:phase_diag_full}.
As we now have incorporated both screening and size dispersity effects,
the resulting behaviour can be seen as a mix of the two individual cases.
Contrast the phase diagram and plotted eigenvalues of Fig.~\ref{fig:phase_diag_full}
to the ones in Fig.~\ref{fig:phase_diag_same_sized}: thanks to the screening effects,
we recover the shifted instability line and the fact that the eigenvalues are null
at $q^2=0$, stopping system-wide instabilities from occurring.
On the other hand, similarly to Fig.~\ref{fig:noscreen_phase_diag}, the eigenvalues can be complex,
but with major differences.
Instead of necessarily having a real positive region, the eigenvalues in Fig.~\ref{fig:phase_diag_full}
can be complex with a positive real part over the whole range of unstable wave vectors,
corresponding to a fully oscillatory instability.
Moreover, in the partially oscillatory regime, the upper eigenvalue exhibits two maxima,
one in the real region and one in the complex region.
whereas in the case without screening it only had a maximum at $q^2=0$.
This implies that, in this regime, the maximally growing mode can be either oscillatory or
non-oscillatory, based on which of these two maxima is the global one.

\subsection{Stoichiometry}
\label{sec:stoich_full_model}

We finally turn to the study of the eigenvectors, more precisely the ratio
$S_{2/1} \equiv \delta \rho_2 / \delta \rho_1$, the sign of which will inform us about the 
tendency of the two species in an unstable binary mixture to aggregate (if positive)
or separate (if negative).
We only study the eigenvector corresponding to the upper eigenvalue $\lambda_+$,
as it is the one driving the instability.
Calculating the eigenvector in \eqref{eq:eig_eq} leads to the expression:
\begin{equation}
    \label{eq:stoich_full_model}
    \begin{aligned}
         S_{2/1} = 
         - \frac{1}{2 \gamma_2} \frac{\gamma_2 \rhz[2]}{\gamma_1 \rhz[1]}
        \bigg[ 
            (\gamma_1 - \gamma_2) + (D_1 - D_2) (d q^2 +\eta) & \\
            + \sqrt{
                \left[ \gamma_1 - \gamma_2 + (D_1 - D_2)(\dqe) \right] ^2
                + 4 \gamma_1 \gamma_2
            }
        \bigg]&
    \end{aligned}
\end{equation}
It can be shown that the term in brackets keeps a constant sign as a function of $q^2$.
By distinguishing between the $\gamma_2 > 0$ and $\gamma_2 < 0$ cases, we can
systematically calculate the sign of $S_{1/2}$ in the regions where the eigenvalue
is real and unstable, leading to the conclusion:
\begin{equation}
    \label{eq:stoich_sign}
    \text{sign} (S_{2/1}) = \text{sign} (\mu_2 / \mu_1)
\end{equation}
Thus, similar to the simple case studied in section~\ref{sec:prl_case}, the sign of the stoichiometry
only depends on the sign of the mobilities, with species having the same mobility
sign tending to aggregate, and species having opposite mobility signs tending to separate.
Note that this result applies to all the cases studied in this paper.
The phase diagrams in Figs.~\ref{fig:phase_diag_same_sized}, \ref{fig:noscreen_phase_diag}
and \ref{fig:phase_diag_full} are then plotted using the same procedure as in
section~\ref{sec:prl_case}: the instability lines are functions of the
$\almurh{m}$, and the stoichiometry depends on the mobilities signs only, so
the phase diagrams can be plotted as a function of $|\alpha_m| \mu_m \rho_{0,m}$ for fixed
signs of the activities.

\section{Discussion}
\label{sec:discus}

In this work, we have explored a general model for the stability of a mixture of active particles
based on the linear stability analysis of continuum equations.
The model studied was a generalization of a simpler model introduced in Ref.~\cite{AgudoCanalejo2019}, in which case the mixture was found to show a system-wide instability if it was overall self-attracting.
We have shown that, if the catalytic activities of the particles have a dependence on the concentration of their substrate,
the interactions become screened, leading to the emergence of a minimum threshold of self-attraction for the instability to occur,
and to the inhibition of system-wide instabilities, which become local (finite wave length).
Accounting for dispersity in the sizes of the active particles, meanwhile, can either lead to the same system-wide instability
observed in the simple model, or to the apparition of an extended, local instability regime with a less strict requirement for the instability.
The existence of size dispersion also allows for the possibility of system-wide, transient oscillations during a global instability.
Finally, combining both screening and size dispersity effects leads to a wide variety of behaviours.
Due to the screening, the instability can only be local, but oscillations are also possible and, depending on the location in phase space, can either represent the dominant unstable mode, or transiently coexist with a more dominant non-oscillatory instability. For each of these cases, we have obtained exact analytical conditions that fully describe the resulting phase diagrams and can be used as guidance in future experimental or simulation studies.
In all the studied cases, the stoichiometry of the growing instability is purely a function of the signs
of the species mobilities, implying that chemotactic species and antichemotactic species tend to separate from each other and aggregate among themselves. 

The instabilities that we predict here at the linear level may also be explored beyond this regime, by means of numerical solution of the continuum equations, or particle-based simulations. Such simulations will allow for the study of the kinetics of the self-organization process, as well as the resulting steady-state configurations of the system.
Of particular interest is the presence or absence of transient oscillations in the instability process.
While our linear stability analysis can yield complex eigenvalues with positive real part, we cannot conclude whether long-lived oscillations will be observed.
In general, the behavior of the system beyond the onset of instability will depend on factors outside the scope of this analysis,
among others the feedback of changes in substrate concentration on the activity and non-linear effects not captured at the linear stability level.
Another limitation of our model is the quasi-static approximation performed for the messenger chemical, implying that we limit
ourselves to cases where the motion of the catalysts is much slower than the diffusion of their substrates and products.

While this paper focused on catalytic particles, many bacteria are known to chemotax in response to chemicals
they themselves secrete or consume, leading to pattern formation \cite{Erban2005,Sengupta2009,Sengupta2011,Meyer2014,Mahdisoltani2021,Gelimson2015}.
In particular, Ref.~\cite{Meyer2014} explores the influence of concentration-dependent chemotactic drift,
stemming from the microscopic characteristics of bacterial receptors, on pattern formation.
Such a concentration-dependent chemotactic mobility could also be incorporated into our model. In turn, our concentration-dependent activity would correspond to cases where the bacteria modulate their production or consumption of chemoattractant or chemorepellant based
on local concentration. Our model could further be applied to bacterial ecosysystems in which different species of bacteria coexist, resulting in inter-species interactions which may be non-reciprocal.

Coming back to the applications at the subcellular level, a natural step for future work is to allow for the enzymes to participate in catalytic cycles, in which the product of one enzyme becomes the substrate of the next enzyme in the cycle \cite{ouazanreboul21}, given that such cycles are ubiquitous in metabolic pathways in the cell. Furthermore, it will be interesting to explore the effect of self-organization on the yield of the associated catalytic reactions. In Ref.~\cite{AgudoCanalejo2019}, it was shown that mixtures of producers and consumers tend to form clusters with just the right stoichiometry that allows for perfect channelling of the chemical released by the producers to be taken up by the consumers in the cluster. The effect of concentration-dependent activities on this phenomenon, as well as the implications on the overall catalytic yield remain to be explored. On the experimental side, a deeper exploration of the dynamics during the formation of metabolons or enzyme-rich condensates may help elucidate whether non-equilibrium chemical-mediated interactions are at play, perhaps in conjunction with passive interactions.

\begin{acknowledgements}
We acknowledge support from the Max Planck School Matter to Life and the MaxSynBio Consortium, which are jointly funded by the Federal Ministry of Education and Research (BMBF) of Germany, and the Max Planck Society.
\end{acknowledgements}

\bibliographystyle{spphys}       
\bibliography{mybib}   

\end{document}